\documentclass[preprint,showpacs,superscriptaddress,groupedaddress,nofootinbib]{revtex4-2}
\usepackage{graphicx}
\usepackage{amssymb}
\usepackage[toc,page]{appendix}
\usepackage{braket}
\newcommand{\del}{\partial}
\newcommand{\f}{\frac}
\usepackage{amsmath}

\begin{document}

\title{Canonical Gravity in Degenerate Limit}
\author{Sandipan Sengupta}
\email{sandipan@phy.iitkgp.ac.in}
\affiliation{Department of Physics, Indian Institute of Technology Kharagpur, Kharagpur-721302, INDIA}

\begin{abstract}

We construct a limit of Hamiltonian gravity as the determinant of the spatial triad (and hence of the four-metric) goes to zero. Within the Barbero-Immirzi $SU(2)$ formulation, we present two possible realizations of this limit, with the consequence that the Hamiltonian constraint becomes simpler and spatial diffeomorphisms become trivial. In the first case, the Hamiltonian constraint exhibits a polynomial structure, being formally similar to the Euclidean Hamitonian constraint of Sen-Ashtekar self-dual formulation. In the latter, the constraints become free from ordering ambiguity. Further, we show that the Carrollian gravity emerges as a special case of this degenerate limit, thus providing it a new geometric interpretation independent of the speed of light or any dimensionful coupling constant ($G$).

\end{abstract}

\maketitle

\newpage


\section{Introduction}

The obstacles to a canonical quantization of Lorentzian gravity are well-known, irrespective of whether the canonical variables are defined within a metric or a connection formulation \cite{kiefer,thiemann}. The structure of the Hamiltonian constraint itself, containing both extrinsic and spatial curvature, poses a serious challenge given the inherent operator ordering ambiguity owing to the appearance of non-commuting operators, non-polynomiality in the canonical variables and regularization issues due to presence of coincident operators or distributions \cite{wdw,thiemann1,al,nicolai}. 
The hunt for a simpler Hamiltonian structure have also led to efforts towards formulating different limits of gravity, with an aim to gain insights exploiting the potential simplifications. This endeavour is reflected in constructions such as the strong coupling limit \cite{teit,henn,pilati, teitelboim}, $G\rightarrow \infty$ limit \cite{husain}, $G\rightarrow 0$ limit \cite{smolin,sa1}, the topological limit \cite{perez} and so on.

In the Hamiltonian theory of Hilbert-Palatini gravity based on the connection formalism, the densitized triad fields emerge as a natural canonical variable (momenta) \cite{kiefer,thiemann,al,peldan}. In a quantum theory, a discrete quantum geometry could be anticipated, where trivial densitized triads could potentially define the lowest eigenstates provided the other (regularized) constraints are also satisfied. Hence, it is worthwhile to explore classical gravity first in a limit where the spatial geometry becomes trivial in the sense of densitized triads having a vanishing determinant. Under certain reasonable assumptions on the lapse function, this implies that the determinant of the four-metric should also vanish in such a limit. Even though discussions on degenerate extensions of general relativity have appeared in the earlier literature \cite{horowitz,jacobson,*romano,kaul2016,*sengupta2016,sengupta2022}, such a limit for gravity has not been explicitly constructed so far.

Analysis of gravity theory in a limit of degenerate metric (tetrad) is also important in order to understand the singularities better. In particular, such a construction could be useful whenever an approach to a singularity is directly associated with a degenerating metric. For instance, such a scenario is realized for spacelike singularities (e.g. cosmological singularity \cite{bkl,bkl1}) where spatial metric determinant approaches a zero. To emphasize, the dynamics resulting from the equations of motion in this limit need not be equivalent to the dynamics far away from singularity. This could provide new insights as to whether gravity behaves differently close to a singularity which is  approached through a limit to degenerate metrics.

To this end, in this note we explore a limit where the determinant of the spatial triad (and also of the four-metric) vanishes. We implement the limit by defining a local scaling transformation on the canonical variables within the connection formalism. 

First, we apply this limit to the Barbero-Immirzi formulation of gravity \cite{barbero,immirzi}.
 We obtain two different reductions of the associated constraints. In both cases, the resulting Hamiltonian theories reflect remarkable simplications. In the first case, we find that the Lorentzian constraints exhibit a formal similarity to the Euclidean constraints of Sen-Ashtekar self-dual gravity \cite{sen,ashtekar,*ashtekar1}, modulo differences regarding the action of spatial diffeomorphism. As an example, we find a solution to this limiting gravity theory in the context of inhomogeneous and anisotropic cosmology. In case of the second limit, we find a Hamiltonian structure whose quantum counterpart would be free of any ordering ambiguity associated with non-commuting operators. Next, we find that a special case of the degenerate limit corresponds to Carrollian gravity \cite{teit,henn,henn1,seng}, the latter being originally defined for a vanishingly small speed of light \cite{ll,nd}. 
 
Apart from exploring possible new limits of Hamiltonian gravity, our analysis could provide a fresh perspective towards the issue of solving the spatial diffeomorphism and the Hamiltonian constraint. To emphasize, we show that the degenerate limit(s) provide a way to solve the spatial diffeomorphism classically, before proceeding towards a canonical quantization. Along with the simpler structure of the Hamiltonian constraint reflected in its polynomial form in particular, achieved without taking recluse to Euclidean gravity, this feature seems promising from the perspective of canonical quantization.

In the next section, we first set up the relevant notation through a very brief outline of Barbero-Immirzi canonical gravity. The limit to a degenerate tetrad through a set of scaling transformations is then introduced. The implications of this limit for the Hamiltonian theory is discussed, along with an explicit example. In Sec-III  we explore a different limit associated with a degenerate tetrad. 
This is followed by an analysis of the special case of global scaling transformations, leading to theories equivalent to Carrollian gravity. We conclude our analysis with a few relevant remarks. Some basic details regarding the symplectic form of Barbero-Immirzi real SU(2) formulation is provided in Appendix A. In Appendix B, as a relevant digression, we present a corollary on Birkhoff's theorem in `electric' Carroll gravity based on the associated Hamiltonian structure.

\section{Degenerate limit of Barbero-Immirzi gravity}

A modern perspective on Hamiltonian gravity based on the connection formulation is provided by the Barbero-Immirzi $SU(2)$ formulation. This is based on a Lagrangian density containing the Hilbert-Palatini and the topological Nieh-Yan density with a constant coefficient \cite{date,kaul}, namely the Barbero-Immirzi parameter :
\begin{eqnarray} \label{L0}
{\cal L}(e,\omega) =  \frac{1}{2\kappa}\Big[e e^{\mu}_I e^{\nu}_{J}R_{\mu\nu}^{~~~ IJ}(\omega)~+~\eta\epsilon^{\mu\nu\alpha\beta}\Big(D_\mu e_\nu^I D_\alpha e_{\beta I}-e_\mu^I e_\nu^J R_{\alpha\beta IJ}(\omega)\Big)\Big] 
\end{eqnarray}
The $SO(3,1)$ field strength above is defined as: $R_{\mu\nu}^{~~~ IJ}(\omega)=\del_{[\mu}\omega_{\nu]}^{~IJ}+\omega_{[\mu}^{~IK}\omega_{\nu]K}^{~~~J}$.
Assuming the standard parametrization of the tetrad as presented in Ref.\cite{kaul,peldan},
the totally constrained Hamiltonian density reads:
\begin{eqnarray}
{\cal H}=\f{1}{2}\omega_t^{~IJ}G_{IJ}+N^a H_a+NH.
\end{eqnarray}
The scalar Lapse $N$, the shift $N^a$ and the temporal connection fields $\omega_t^{~IJ}$ emerge as Lagrange multipliers above, whereas $H\approx 0,~H_a \approx 0,~G_{IJ}\equiv (G_{0i},~G_{ij})\approx 0$ denote the Hamiltonian, spatial diffeomorphism, boost and rotation constraints, respectively. Their explicit expressions are presented later. The spatial metric is defined as: $ q_{ab} ~ := ~e_{a}^{I}e_{bI}$, and the tetrad determinant as: $e := det(e^{I}_{\mu}) = N\sqrt{q}$. Further relevant details in this context are provided in Appendix A.


In time gauge ($e_a^0\approx 0$), which fixes the boost freedom, the number of canonical pairs get reduced and the expressions of the constraints simplify. The resulting symplectric form reads \footnote{The original canonical pair is defined as: $\omega_a^{(\eta)IJ}:=\omega_a^{IJ}+\frac{\eta}{2}\epsilon^{IJKL}\omega_{aKL},~\pi^a_{~IJ}:=\frac{1}{2\kappa}\epsilon^{abc}\epsilon_{IJKL}e_b^K e_c^L$}:
\begin{eqnarray}
\Omega=\frac{1}{2}\pi^a_{~IJ}\del_t \omega_a^{(\eta)IJ}=E^a_{i}\del_t A_a^i,
\end{eqnarray}
where the conjugate pair is given by: $A_a^i:=\omega_a^{0i}+\frac{\eta}{2}\epsilon^{ijk}\omega_a^{jk},~E^a_i:=\pi^a_{~0i}=e e^a_i,~q=\det E^a_i:=E$.
The connection fields $A_a^i$ emerge as the $SU(2)$ gauge field. The momenta $E^a_i$ are known as the densitized spatial triad fields. These conjugate fields are analogous to the gauge potential and the electric fields of $SU(2)$ Yang-Mills theory, respectively. 

In terms of this canonical pair in the symplectic form, the set of constraints read \cite{sa,date}:
\begin{eqnarray}\label{bi-constr}
G^{rot}_i~&=&~-\eta D_a(A) E^a_i\approx 0,\nonumber\\
H_a~&=&~E^b_i F_{ab}^i(A)+\Big(\eta+\frac{1}{\eta}\Big)\Bigg(A_a^i+\frac{\eta}{2}\epsilon^{ijk}\omega_a^{~jk}(E)\Bigg)G^{rot}_i\approx 0,\nonumber\\
H~&=&~\frac{\kappa}{2\sqrt{E}}E^{a}_i E^{b}_j \Big[\eta\epsilon^{ijk} F_{ab}^k (A)
-(1+\eta^2)\Big(\del_{[a}\omega_{b]}^{~ij}(E)+\omega_{[a}^{~ik}(E)\omega_{b]}^{~kj}(E)\Big)\Big]\approx 0,
\end{eqnarray}
where $D_a (A) V^i:=\del_a V^i-\frac{1}{\eta}\epsilon^{ijk}A_a^j V^k$ and $F_{ab}^{~i}(A):=\del_{[a}A_{b]}^i-\frac{1}{\eta}\epsilon^{ijk}A_a^j A_b^k$ denote the $SU(2)$ covariant derivative and field-strength, respectively, and the spatial connection components $\omega_a^{~ij}(E)$ are defined as:
\begin{eqnarray}\label{omega}
\omega_a^{~ij}(E)~=~\frac{1}{2}E_a^{[i}\del_b E^{bj]}~+~\frac{1}{2}\epsilon^{ijk}E_a^l\Big[\epsilon^{mn(k}E^{l)}_b-\epsilon^{mnp}E_b^p \delta^{kl}\Big]\del_c\big(E^c_m E^b_n\big)
\end{eqnarray}
These three constraints above are the generators of the symmetry transformations associated with the  spatial rotation, spatial diffeomorphisms (modulo rotations) and temporal translations, respectively.  

\subsection*{A limit to degenerate tetrad}

In order to implement the limit to vanishing triad determinant, we introduce a local scaling transformation on the triad fields as:
 \begin{eqnarray}\label{e-scaling}
 e_a^i=\delta_{(a)}e^{'i}_a.
 \end{eqnarray}
 The brackets in the spatial label `$(a)$', which can run over the three space coordinates, imply that it is not summed over when repeated. The primed variables are assumed to remain finite in the limit.  To emphasize, each of the three vectors for a fixed $(a)$ need not vanish in this limit.  The triad determinant $\sqrt{q}=\delta \sqrt{q'}$ goes to zero as $\delta=\prod_a \delta_{(a)}\rightarrow 0$, which defines the `degenerate' limit. This in turn implies that:
 $\delta_{(a)}\delta_{(b)}=\frac{\delta}{\delta_{(c)}}\rightarrow 0$ for any $a\neq b \neq c$ in the limit. An explicit realization of these properties is presented through an example later.
 
Under the condition that the symplectic form be preserved, eq.(\ref{e-scaling}) leads to the following transformation of the canonical pairs:
\begin{eqnarray}\label{sclaw}
A_a^{i}&=&\frac{\delta_{(a)}}{\delta}A_a^{'i},~E^a_i=\frac{\delta}{\delta_{(a)}}E^{'a}_i.
\end{eqnarray}
Note that $E^a_i\rightarrow 0$ as $\delta\rightarrow 0$. The natural transformation of the spatial derivative which preserves the tetrad one-form is given by: $\del'_a=\frac{\del_a}{\delta_{(a)}}$. Throughout this article we assume that the coupling constants do not scale.

Under (\ref{sclaw}), the transformed constraints are given by:
\begin{eqnarray}
G^{rot}_i &=&\epsilon^{ijk}A_a^{'k}E^{'a}_j-\eta\frac{\delta}{\delta_{(a)}}\del_a E^{'a}_i-\eta E^{'a}_i \del_a \frac{\delta}{\delta_{(a)}};\nonumber\\
H_a &=&\frac{\delta_{(a)}}{\delta}\Big[-\frac{1}{\eta} \epsilon^{ijk}A_a^{'j}A_{b}^{'k}E^{'b}_i-\frac{\delta}{\delta_{(a)}} E^{'b}_i \del_{a}A_{b}^{'i}
-\frac{\delta}{\delta_{(b)}}E^{'b}_i \del_{b}A_{a}^{'i} \nonumber\\
&-& E^{'b}_i A_b^{'i} \frac{\delta}{\delta_{(a)}}\frac{\delta_{(b)}}{\delta}\del_a \frac{\delta}{\delta_{(b)}}+E^{'b}_i A_b^{'i}\frac{\delta}{\delta_{(b)}}\frac{\delta_{(a)}}{\delta}\del_a \frac{\delta}{\delta_{(a)}}
+ \Big(\eta+\frac{1}{\eta}\Big)\Bigg(A_a^{'i}+\frac{\eta}{2}\epsilon^{ijk}\omega_a^{'jk}\Bigg)G^{'rot}_i\Big]\nonumber\\
H &=&\frac{\kappa}{2\delta\sqrt{E'}}E^{'[a}_i E^{'b]}_j \Big[A_a^{'i} A_b^{'j}~+~\delta^2 \Big(\frac{1}{\delta_{(a)}}\del_a \omega_b^{'ij}+\omega_a^{'ik}\omega_b^{'kj}\Big)+\omega^{'ij}_b\frac{\delta^2}{\delta_{(a)} \delta_{(b)}}\del_a \frac{\delta_{(b)}}{\delta}\Big]
\end{eqnarray}
In writing the last line above we have used the expression (\ref{omega}) for the spatial connection components.
With the assumptions: $\del_a\frac{\delta}{\delta_{(b)}}\rightarrow 0\Rightarrow ~\frac{\delta}{\delta_{(a)}}\frac{\delta}{\delta_{(b)}}\del_c \frac{\delta_{(b)}}{\delta}\rightarrow 0$ for any $a,b,c$, the spatial derivative terms in the constraints above drop out. Redefining the shift and lapse as: $N^a=\frac{\delta}{\delta_{(a)}}N^{'a},~N=\delta N$, we find, as $\delta\rightarrow 0$:
\begin{eqnarray}\label{ham}
G^{'rot}_i &=&\epsilon^{ijk}A_a^{'k}E^{'a}_j;\nonumber\\
H'_a &=&\eta A_a^{'i}G^{'rot}_i;\nonumber\\
H' &=&-\frac{\kappa}{2\sqrt{E'}}E^{'[a}_i E^{'b]}_j A_a^{'i} A_b^{'j}
\end{eqnarray}

Noting that the $SU(2)$ covariant derivative and the field-strength in the limit are given by: $D'_a (A') E^{'a}_i=-\frac{1}{\eta}\epsilon^{ijk} A_a^{'j}E^{'ak},~F_{ab}^{'~i}= -\frac{1}{\eta}\epsilon^{ijk}A^{'j}_a  A_b^{'k}$, the above set may be rewritten as:
\begin{eqnarray}\label{ham1}
G^{'rot}_i &=&\eta D'_a(A')E^{'a}_i;\nonumber\\
H'_a &=&E^{'b}_i F_{ab}^{'~i}(A');\nonumber\\
H' &=&\frac{\kappa \eta}{2\sqrt{E'}}\epsilon^{ijk}E^{'a}_i E^{'b}_j F_{ab}^{'~k}(A')
\end{eqnarray}
Note that modulo an overall density factor which could be absorbed away through a densitized lapse function, the Hamiltonian constraint has now become a quadratic polynomial in the canonical variables. 

Let us observe that this same feature arises in the 
Sen-Ashtekar self-dual connection formulation, which, however, corresponds to Euclidean gravity (and could be obtained by setting $\eta=i$ in Barbero-Immirzi formulation). The other important difference is that the Lorentzian set here corresponds to a simpler (and weaker) canonical structure, given that the spatial diffeomorphism constraints are trivially satisfied.

It is also straightforward to observe that the reduced constraints preserve the $SU(2)$ interpretation. For instance, the rotation constraint generates the following transformation of the canonical pair:
\begin{eqnarray}
&&\delta A_a^i=[A_a^i,\int \theta^k G_k^{rot}]=\epsilon^{ijk}A_a^j \theta^k=\eta D_a\theta^i,\nonumber\\
&&\delta E^a_i=[E^a_i,\int \theta^k G_k^{rot}]=\epsilon^{ijk}\theta^k E^a_j
\end{eqnarray}
using the fact that in the limit the gauge-covariant derivative reduces to: $D_a\theta^i=\frac{1}{\eta}\epsilon^{ijk}A_a^j \theta^k$. These are manifestly same as the standard $SU(2)$ gauge transformations acting on the gauge potential and the electric field components, respectively. 

\subsection*{Example: Ultralocal physics}

The properties of the scaling fields $\delta_{(a)}$ allow us to describe a dynamical spacetime where all the spatial directions need not contract in general as the triad (tetrad) degenerates.
As an example of an inhomogeneous and anisotropic universe, let us consider the following spacetime:
\begin{eqnarray*}
ds^2=-dt^2+t^{2\alpha(x,y,z)}dx^2+t^{2\beta(x,y,z)}dy^2+t^{2\gamma(x,y,z)}dz^2
\end{eqnarray*}
where we assume that: $\alpha+\beta>0, ~\beta+\gamma>0,~\alpha+\gamma>0$. A similar metric was first considered by BKL \cite{bkl,*bkl1}, which could be a solution of  Einstein gravity in presence of matter. Assuming that the effect of matter could be neglected in the degenerate limit (which here represents a curvature singularity at $t=0$), we proceed to find the associated solution. 

Let us first demonstrate that the assumed generic properties of the scaling transformation are all satisfied. The spatial metric (as well as the four-metric) degenerates as $t\rightarrow 0$. This limit could be parametrized as $\epsilon\rightarrow 0$ where $t=\epsilon t'$ ($t'$ being finite). Inserting this in the metric solution just obtained above, we find the components of $\delta_{(a)}$ defined in eq.(\ref{e-scaling}) as:
\begin{eqnarray*}
\delta_{(x)}=\epsilon^{\alpha(x,y,z)},~\delta_{(y)}=\epsilon^{\beta(x,y,z)},~\delta_{(z)}=\epsilon^{\gamma(x,y,z)}.
\end{eqnarray*}
Thus, the degenerate limit is equivalent to $\delta=\epsilon^{\alpha+\beta+\delta}\rightarrow 0$.
Let us note that since $\epsilon^\sigma \ln\epsilon\rightarrow 0$ as $\epsilon \rightarrow 0$ for any $\sigma>0$ (as a consequence of L'Hospital's rule), we have: $\del_a \frac{\delta}{\delta_{(b)}}\rightarrow 0$ for any $a,b$.
This in turn implies that $\frac{\delta}{\delta_{(a)}}\frac{\delta_{(b)}}{\delta}\del_c \frac{\delta}{\delta_{(b)}}\rightarrow 0$ for any $a,b,c$.
Thus, the scaling transformation satisfies the assumptions made earlier.

The nontrivial components of the momenta are given by (suppressing the primes which are implicit in the variables and constraints):
\begin{eqnarray*}
E^x_1=t^{\beta+\gamma},~E^y_2=t^{\gamma+\alpha},~E^z_3=t^{\alpha+\beta}.
\end{eqnarray*}
The time evolution of the canonical pair with respect to the Hamiltonian constraint (\ref{ham}) leads to (in the gauge $\omega_t^{~ij}=0$):
\begin{eqnarray}\label{t-evol}
\dot{E}^a_i&=&\left[E^a_i,\int {\cal H}\right]=\frac{N}{\sqrt{E}}E^{[a}_i E^{b]}_j A_b^j;\nonumber\\
\dot{A}_a^i&=&\left[A_a^i,\int {\cal H}\right]=-\frac{N}{\sqrt{E}}\Big[E^b_j A_{[a}^i A_{b]}^j+\frac{\sqrt{E}}{2}E_a^i H\Big]
\end{eqnarray}
From the first equation above, the nontrivial components of the coordinates are obtained as:
\begin{eqnarray*}
 A_x^{1}=\alpha t^{\alpha -1},~A_y^{2}=\beta t^{\beta-1},~A_z^{3}=\gamma t^{\gamma-1}
\end{eqnarray*}
The second evolution equation in (\ref{t-evol}) for the coordinates, along with the Hamiltonian constraint,  imply:
\begin{eqnarray}\label{eqg}
&&\alpha\beta+\beta\gamma+\gamma \alpha=0, \nonumber\\
&&\alpha(\alpha+\beta+\gamma-1)=0,\nonumber\\
&&\beta(\alpha+\beta+\gamma-1)=0,\nonumber\\
&&\gamma(\alpha+\beta+\gamma-1)=0.
\end{eqnarray}
Assuming $\alpha\neq 0,~\beta\neq 0,~\gamma\neq 0$, these are solved as: $\alpha+\beta+\gamma=1=\alpha^2+\beta^2+\gamma^2$.
The rotation constraints are trivially satisfied. 

As the above solution reflects, the spatial dependence of these exponents are arbitrary, unlike in the case of Kasner solution of vacuum Einstein gravity where these are forced to be constants by the field equations. Thus, the spatial points are uncorrelated and at each point there could be an independent set of Kasner exponents. In other words, the degenerate limit in this context leads to a classical realization of ultralocal physics.

\section{An alternative limit: Hamiltonian structure without ordering ambiguity}

In this section, we explore a different scaling transformation to set up the degenerate limit. This is to elucidate the fact that it is possible to define the limit so that the spatial derivative terms do not necessarily disappear from the constraints.

Let us consider the following scaling transformations:
\begin{eqnarray}\label{sclaw-diff}
A_a^{i}&=&\delta_{(a)}A_a^{'i},~E^a_i=\frac{\delta}{\delta_{(a)}}E^{'a}_i,
\end{eqnarray}
To compare with eq.(\ref{sclaw}), the triad fields transform as earlier, whereas the gauge fields now transform differently. Note that this does not preserve the symplectic form. 
Let us also demand that $(E^a_i \del_a)$ be preserved (instead of the tetrad one-form), implying that: $\del_a=\frac{\delta_{(a)}}{\delta}\del'_a$.

Using the above transformations, the constraints become:
\begin{eqnarray}
G^{rot}_i &=&-\eta\frac{\delta}{\delta_{(a)}}\del_a E^{'a}_i+\delta\epsilon^{ijk}A_a^{'k}E^{'a}_j-\eta E^{'a}_i \del_a \frac{\delta}{\delta_{(a)}},\nonumber\\
H_a &=&\frac{\delta_{(a)}}{\delta}\Big[-\frac{\delta}{\eta} \epsilon^{ijk}A_a^{'j}A_{b}^{'k}E^{'b}_i-\frac{\delta^2}{\delta_{(a)}} E^{'b}_i \del_{a}A_{b}^{'i}
- \frac{\delta^2}{\delta_{(b)}}E^{'b}_i \del_{b}A_{a}^{'i} \nonumber\\
&+& E^{'b}_i A_b^{'i} \frac{\delta}{\delta_{(a)}}\frac{\delta}{\delta_{(b)}}\del_a \delta_{(b)}- E^{'b}_i A_a^{'i}\frac{\delta}{\delta_{(b)}}\frac{\delta}{\delta_{(a)}}\del_b \delta_{(a)}+ \Big(\eta+\frac{1}{\eta}\Big)\delta \Bigg(A_a^{'i}+\frac{\eta}{2}\epsilon^{ijk}\omega_a^{'jk}\Bigg)G^{'rot}_i\Big]
,\nonumber\\
H &=& \frac{\kappa }{2\sqrt{E'}}E^{'[a}_i E^{'b]}_j\Big[A_a^{'i} A_b^{'j}~+~(1+\eta^2) \delta^2\Big(\frac{1}{\delta_{(a)}}\del_a \omega_b^{'ij}+\omega_a^{'ik}\omega_b^{'kj}\Big)~+~(1+\eta^2)\omega^{'ij}_b\frac{\delta^2}{\delta_{(a)} \delta_{(b)}}\del_a \frac{\delta_{(b)}}{\delta}\Big]\nonumber\\
&&~
\end{eqnarray}
The properties of the scaling transformation imply that the terms involving derivatives of the scaling fields vanish in the limit, leading to:
\begin{eqnarray}
G^{'rot}_i &=&-\eta D'_a E^{'a}_i,\nonumber\\
H'_a &=& 0,\nonumber\\
H' &=& -\frac{\kappa (1+\eta^2)}{2\sqrt{E'}}E^{'a}_i E^{'b}_j \bar{R}_{ab}^{'ij}(E'),
\end{eqnarray}
where $D'_a E^{'a}_i= \del'_a E^{'a}_i$ and $\bar{R}_{ab}^{'ij}(E')=\del_{[a} \omega_{b]}^{'ij}(E')+\omega_{[a}^{'ik}(E')\omega_{b]}^{'kj}(E')$ is the spatial three-curvature. In obtaining the above, we have redefined the Lagrange multipliers as: $\omega_t^{~ij}=\frac{1}{\delta}\omega_t^{'ij},~N^a=\frac{\delta}{\delta_{(a)}}N^{'a}$.

This reflects another intriguing limit, where the diffeomorphism becomes trivial as in the earlier section. However, the rotation and Hamiltonian constraints above depend only on the conjugate momenta $E^a_i$. Since these have no variables with nontrivial Poisson brackets among themselves, the quantum constraints would have no non-commuting operators. In other words, the quantization of this theory would be free of any operator ordering ambiguity problem, even though regularization issues associated with product of commuting operators  would still be alive. 

\section{Degenerate limit and Carroll gravity}

In this section, we demonstrate that the degenerate limit of Hilbert-Palatini gravity reproduce Carroll gravity when the scaling transformations parametrizing the vanishing of the determinant are made global: $\delta_{(a)}\equiv const$. 
Carroll gravity had originally been formulated as a truncated Hamiltonian theory in the limit $c\rightarrow 0$ \cite{henn1}. Recently, both the `electric' and `magnetic' forms of this formulation was shown to originate from the Hilbert-Palatini action corresponding to first-order gravity, and the associated connection-based Hamiltonian formulations were also presented \cite{seng}. 

Using the same parametrization of the variables as earlier, the Hilbert-Palatini constraints are given by:
\begin{eqnarray}\label{HPconstraints}
G^{rot}_i~&=&~\epsilon^{ijk}K_a^k E^a_j,\nonumber\\
H_a~&=&~E^b_i\del_{[a} K_{b]}^i-K_a^i \del_b E^b_i,\nonumber\\
H~&=&~-~\frac{\kappa}{2\sqrt{E}}E^{[a}_i E^{b]}_j \Big[K_a^i K_b^j~+~\del_a \omega_b^{~ij}(E)+\omega_a^{~ik}(E)\omega_b^{~kj}(E)\Big]
\end{eqnarray}
where $K_a^i:= \omega_a^{~0i}$. 
We now proceed to find the reduced set of constraints in the degenerate limit implemented through a global scaling transformation.

\subsection{Electric Carroll gravity}
First we consider the analogue of the scaling transformation (\ref{sclaw}) on conjugate pairs is given by:
\begin{eqnarray}\label{sclaw1}
K_a^{i}&=&\frac{\delta_{(a)}}{\delta}K_a^{'i},~E^a_i=\frac{\delta}{\delta_{(a)}}E^{'a}_i
\end{eqnarray}
Note that the transformation of the derivatives and of $\omega_a^{'ij}(E)$ remain the same as in section-II.
Under (\ref{sclaw1}), the rotation constraint under the scaling remain form-invariant:
 \begin{eqnarray}\label{rotc}
 G^{rot}_i=\epsilon^{ijk}K_a^{'k}E^{'a}_j:=G^{'rot}_i
 \end{eqnarray}
The spatial diffeomorphism constraints transform as:
\begin{eqnarray*}
H_a=\delta_{(a)}\Big[\frac{1}{\delta_{(a)}}E^{'b}_i \del_{a}K_{b}^{'i}-\frac{1}{\delta_{(b)}}E^{'b}_i \del_{b}K_{a}^{'i}-\frac{1}{\delta_{(b)}}K_a^{'i} \del_b E^{'b}_i \Big]
\end{eqnarray*}
Redefining the shift as: $N^a:=\frac{1}{\delta_{(a)}}N^{'a}$, we find:
\begin{eqnarray}\label{diffc}
H'_a=E^{'b}_i \del'_{a}K_{b}^{'i}-E^{'b}_i \del'_{b}K_{a}^{'i}-K_a^{'i} \del'_b E^{'b}_i 
\end{eqnarray}
Finally, the Hamiltonian constraint becomes:
\begin{eqnarray*}
H=-\frac{\kappa}{2\delta\sqrt{E'}}E^{'[a}_i E^{'b]}_j \Big[K_a^{'i} K_b^{'j}+~\delta^2 \Big(\frac{1}{\delta_{(a)}}\del_a \omega_b^{'ij}+\omega_a^{'ik}\omega_b^{'kj}\Big)\Big]
\end{eqnarray*}
As $\delta\rightarrow 0$, we obtain:
\begin{eqnarray}\label{hc}
H'=-\frac{\kappa}{2\sqrt{E'}}E^{'[a}_i E^{'b]}_j K_a^{'i} K_b^{'j}
\end{eqnarray}
with the redefinition: $N:=\delta N'$.

Notice that the Hamiltonian structure above corresponds precisely to the `electric' Carroll limit in connection formulation \cite{seng}. The additional factor $\frac{1}{\sqrt{E'}}$ in the Hamiltonian constraint $H$ here compared to ref.\cite{seng} originates due to the use of a scalar lapse as opposed to a densitized one. 

 
\subsection{Magnetic Carroll gravity}

Next, we consider a global scaling analogous to eq.(\ref{sclaw-diff}), which provides an alternative realization of the degenerate limit:
\begin{eqnarray}
K_a^{i}&=&\delta_{(a)}K_a^{'i},~E^a_i=\frac{\delta}{\delta_{(a)}}E^{'a}_i.
\end{eqnarray}
Following the procedure already elucidated in section-III, we find:
 \begin{eqnarray*}
 && G^{rot}_i=\delta \epsilon^{ijk}K_a^{'k}E^{'a}_j,\nonumber\\
&&H_a=\delta_{(a)}\Big[\frac{\delta}{\delta_{(a)}}E^{'b}_i \del_{a}K_{b}^{'i}-\frac{\delta}{\delta_{(b)}}E^{'b}_i \del_{b}K_{a}^{'i}-\frac{\delta}{\delta_{(b)}}K_a^{'i} \del_b E^{'b}_i \Big]\nonumber\\
&& H=-\frac{\kappa}{2\delta\sqrt{E'}}E^{'[a}_i E^{'b]}_j \Big[\frac{\delta}{\delta_{(a)}}\del_a \omega_b^{'ij}(E')+\omega_a^{'ik}(E')\omega_b^{'kj}(E')~+~\delta^2 K_a^{'i} K_b^{'j}\Big]
\end{eqnarray*}
Using the redefinition of the Lagrange multipliers as: $\omega_t^{~ij}=\frac{1}{\delta}\omega_t^{'ij},~N^a=\frac{1}{\delta_{(a)}}N^{'a},~N=\delta N'$, we find that the rotation and spatial diffeomorphism constraints remain form-invariant, whereas the Hamiltonian constraint reduces to:
\begin{eqnarray}\label{H}
H'&=&-\frac{\kappa}{2\sqrt{E'}}E^{'[a}_i E^{'b]}_j \Big[\del'_a \omega_b^{'ij}(E')+\omega_a^{'ik}(E')\omega_b^{'kj}(E')\Big]
\end{eqnarray}
This set of constraints resembles the Hamiltonian form of the `magnetic' Carroll gravity \cite{berg,henn1} in the connection formulation \cite{seng}. 

Note that the `electric' and `magnetic' Carroll limits could be obtained through a $c\rightarrow 0$ limit, which could in turn be defined through a set of global transformations on the tetrad and connection variables leading to a singular determinant, as explicitly demonstrated in Ref.\cite{seng}. While the transformations here are global as well, they act  on the canonical variables in a different way. 

\section{Conclusions}

Here we have presented a Hamiltonian form of gravity theory in the limit of a  vanishing triad determinant. This is equivalent to the limit to a vanishing metric determinant. This limit is relevant in the context of connection-triad formulation of canonical quantum gravity. Cases of spacelike singularities (e.g. the cosmological singularity) provide a more specific context where such a limit typically shows up.

We have explored the implications of the possible degenerate limits within the Barbero-Immirzi formulation, which provides a modern perspective to canonical gravity. We elucidate two possible realizations, and find that in both cases the constraint structure is significantly simplified compared to standard Hilbert-Palatini gravity. In particular, the spatial diffeomorphism constraints are shown to be solved trivially, while the Hamiltonian constraint exhibits intriguing simplifications. In the first case, the Hamiltonian constraint becomes a polynomial function of the canonical variables (upto an overall density factor), and in the second it becomes free of ordering ambiguity while being non-polynomial. Either cases provide simpler starting points for a canonical quantization programme when compared to the Hilbert-Palatini Hamiltonian theory. To emphasize, our analysis leads to two main insights regarding the open questions involving the diffeomorphism and Hamiltonian constraints within a canonical quantization approach. Firstly, we learn that the spatial diffeomorphism constraints could be satisfied classically in any physical scenario (e.g. as a spacelike singularity is approached) where the spatial metric tends to degenerate. Hence, it might not be necessary in this regime to carry the diffeomorphisms all the way to the quantum theory, invoke its operator representation and then find its solutions unlike the usual approach. Secondly, the simplification of the Hamiltonian constraint is not an exclusive feature of Ashtekar's self-dual gravity with complex variables or of Carrollian gravity typically interpreted as the low speed of light limit. 

We have also shown that in the first of the two degenerate limits, an ultralocal behaviour (absence of space derivatives) emerges. While a similar behaviour is encountered in  some other contexts, for instance, in the context of the BKL limit to a cosmological singularity, let us emphasize that the associated Hamiltonian structure here is quite different from the BKL Hamiltonian formulation \cite{sloan,*sloan1}. This earlier work is built upon a set of densitized spatial derivatives, whose action is defined in such a way that leads to a set of truncated constraints. Unlike our case, this set contains the spatial diffeomorphism constraints. We also note that ultralocality has been reported in the earlier literature in contexts such as electric Carrollian gravity \cite{henn} or self-dual Euclidean gravity in the strong coupling limit \cite{husain}.  

Classically, the fact that the set of constraints in the degenerate limit is weaker suggests that the associated ultralocal behaviour is potentially associated with a higher number of degrees of freedom. The true nature and role of these additional degrees of freedom need a deeper study.

Regarding the possible method to proceed to a canonical quantization, the degenerate limit in Sec-II could be taken up based on either the triad or connection representation. For the alternative limit discussed in Sec-III, a triad representation would be more appropriate since all the constraints in this case depend only on the densitized triad and its derivatives. It might also be worthwhile to explore if  a perturbative theory of gravity could be defined around the degenerate limit. 


Finally, our analysis provides a different interpretation to Carroll gravity in terms of the spatial metric determinant. Note that the original Carrollian formulations defined in terms of dimensionful constants, i.e. the $c\rightarrow 0$ or $G\rightarrow \infty$ limit, are severely dependent on the choice of units. The framework here is manifestly free of such ambiguities. Further, given the connection between Carrollian physics and null hypersurfaces (e.g. black hole event horizons) \cite{bagchi}, it would be interesting to explore a potential role of a degenerate metric limit in such contexts. As our results in Appendix B reflect, the spherically symmetric sector admits solutions which evade the Birkhoff's theorem in the degenerate limit (as well as in electric Carroll gravity). It could be worthwhile to understand the implications in the broad context of black hole singularities.

\begin{acknowledgments}
This work is supported (in part) by the MATRICS project grant MTR/2021/000008, SERB, 
Govt. of India. 

\end{acknowledgments}

\bibliography{bi-limit}

\appendix

\section{Derivation of the symplectic form}
We begin with the standard parametrization of the tetrad fields in terms of the lapse function and shift vector \cite{peldan, sa, kaul}:
\begin{eqnarray*}
&& e^{I}_{t} =  NM^I+N^a V_{a}^{I} , ~ e^{I}_{a} = 
V^{I}_{a};\nonumber\\
&& M_{I} V_{a}^{I} = 0 , ~ M_{I} M^{I} =  -1;\nonumber\\
&& e^{t}_{I}  =  -\frac{M_{I}}{N} , ~ e^{a}_{I} ~ = ~
V^{a}_{I}+\frac{N^{a} M_{I}}{N}  ~ ~; \nonumber \\
&& M^{I}V_{I}^{a} :=  0 ,~ V_a^I V^b_I ~ := ~ \delta_a^b ,~ V_a^I
V^a_J:= \delta^I_J + M^I M_J ~~;\nonumber\\
&& q:=\det q_{ab}=\det(V_a^I V_{bI})
\end{eqnarray*} 
Using this decomposition, the Lagrangian density could be rewritten as:
\begin{eqnarray} \label{ny}
{\cal L}(e,\omega) &=&  \frac{1}{2\kappa}\Big[e e^{\mu}_I e^{\nu}_{J}R_{\mu\nu}^{~~~ IJ}(\omega)~+~\eta\epsilon^{\mu\nu\alpha\beta}\Big(D_\mu e_\nu^I D_\alpha e_{\beta I}-e_\mu^I e_\nu^J R_{\alpha\beta IJ}(\omega)\Big)\Big] \nonumber\\
&=&\frac{1}{2}\pi^a_{~IJ}\del_t\omega_a^{(\eta)IJ}-\f{1}{2}\omega_t^{~IJ}G_{IJ}-N^a H_a-NH
\end{eqnarray}
where the momenta conjugate to the canonical coordinates $\omega_a^{(\eta)IJ}$ are defined as: $\pi^a_{~IJ}=ee^{t}_{[I}e^{a}_{J]}$. Since the variables $\omega_t^{~IJ},~N^a$ and $N$ have no conjugate momenta, these lead to secondary constraints given by the following expressions, respectively:
\begin{eqnarray}
G_{IJ}&=&-D_a \pi^{(\eta)a}_{IJ}a\approx 0,\nonumber\\
H_a&=&\frac{1}{2}\pi^b_{~IJ}R^{(\eta)IJ}_{ab}\approx 0,\nonumber\\
H&=&\frac{1}{2\sqrt{q}}\pi^a_{~IK}\pi^b_{~JL}\eta^{KL} R^{(\eta)IJ}_{ab}\approx 0,
\end{eqnarray} 
where $\eta^{IJ}$ is the internal Minkowski metric. From eq.(\ref{ny}), the symplectic form could be read off as:
$\Omega=\frac{1}{2}\pi^a_{~IJ}\del_t\omega_a^{(\eta)IJ}$. 

The original canonical momenta $\pi^a_{~IJ}$ are not all independent, as these are associated with the (set of six) primary constraints:
\begin{eqnarray}
\epsilon^{IJKL}\pi^a_{~IJ}\pi^b_{~KL}\approx 0
\end{eqnarray}
These constraints are solved \cite{sa} by the following choice of variables, eliminating six redundant components out of the nine components of $\pi^a_{~ij}$, leaving only three independent ones ($\chi_i$):
\begin{eqnarray}
\pi^a_{~ij}=\chi_{[i}\pi^a_{~0j]}
\end{eqnarray}
The symplectric form becomes:
\begin{eqnarray}\label{omega}
\Omega=\frac{1}{2}\pi^a_{~IJ}\del_t \omega_a^{(\eta)IJ}=\pi^a_{~0i}\del_t\omega_a^{(\eta)0i}+\frac{1}{2}\pi^a_{~ij}\del_t\omega_a^{(\eta)ij}=E^a_{i}\del_t A_a^i+\zeta^i\del_t \chi_i
\end{eqnarray}
where we have introduced the new variables as: $E^a_i:=\pi^a_{~0i},~A_a^i:=\omega_a^{(\eta)0i}-\chi_j \omega_a^{(\eta)ij},\zeta^i=-\pi^a_j\omega_a^{(\eta)ij}$.
Note that $q=\det E^a_i:=E$. 
In the above, the L.H.S. contains eighteen canonical pairs, whereas the R.H.S. has only twelve. Thus, the six redundant components of $\omega_a^{~ij}$ have disappeared. 

Next, in the time-gauge, $\chi_i=0$. With this, the symplectic form (\ref{omega}) finally reduces to:
\begin{eqnarray}
\Omega=E^a_{i}\del_t A_a^i
\end{eqnarray}
This leaves $(A_a^i,E^b_j)$ as the only independent canonical pair subject to the constraints.

\section{A Corollary on the violation of Birkhoff's theorem in electric Carroll gravity}

As a relevant extension to our discussion of Carroll gravity in this general context, we shall elucidate an important implication of the limiting Hamiltonian theory in the context of spherical symmetry.

Let us first consider `electric' Carrollian limit of gravity defined by the constraints (\ref{rotc}), (\ref{diffc}) and (\ref{hc}).

Given the standard ansatz for a spherically symmetric spacetime:
\begin{eqnarray*}\label{sph}
ds^2=-e^{\mu(t,r)} dt^2+e^{\lambda(t,r)}dr^2+r^2(d\theta^2+\sin^2\theta d\phi^2),
\end{eqnarray*}
the canonical momenta $E^a_i$ are given by:
\begin{eqnarray}\label{E}
E^r_1=r^2\sin\theta,~E^\theta_2=e^{\frac{\lambda}{2}}r\sin\theta,~E^\phi_3=e^{\frac{\lambda}{2}}r.
\end{eqnarray}
From their time evolution with respect to the total Hamiltonian, the only nontrivial canonical coordinate reads:
\begin{eqnarray}
K_r^1:=\omega_r^{~01}=\frac{1}{2}\dot{\lambda}e^{\frac{\lambda-\mu}{2}}.
\end{eqnarray}
All the constraints are satisfied trivially, except the radial component of the spatial diffeomorphism, which implies:
\begin{eqnarray}
Q_r^1=0\Rightarrow \lambda=\lambda(r)
\end{eqnarray}
Hence, $\lambda$ could be any arbitrary radial function, whereas the lapse (or, $\mu(t,r)$) is left undetermined. 

This demonstrates that the Birkhoff's theorem is violated by the `electric' limit of Carrollian gravity. 

For the other possible Carrollian case, namely the so-called `magnetic' limit, it is straightforward to show that the limiting theory reproduces the Schwarzschild geometry as the unique solution, thus satisfying the Birkhoff's theorem.

\end{document}